\documentclass[prl,twocolumn,showpacs,amsmath,amssymb,superscriptaddress]{revtex4}
\usepackage{graphicx}
\usepackage{dcolumn}
\usepackage{bm}
\usepackage{epsfig}

\begin{document}

%\preprint{APS/123-QED}

\title{Anomalous temperature evolution of the internal magnetic field 
distribution in the charge-ordered triangular antiferromagnet AgNiO$_{2}$}

\author{T. Lancaster}
\email{t.lancaster1@physics.ox.ac.uk} 
\author{S.J. Blundell} 
\author{P.J. Baker}
\author{M.L. Brooks} 
\author{W. Hayes}
\affiliation{
Clarendon Laboratory, Oxford University Department of Physics, Parks
Road, Oxford, OX1 3PU, UK}
\author{F.L. Pratt}
\affiliation{
ISIS Facility, Rutherford Appleton Laboratory, Chilton, Oxfordshire 
OX11 0QX, UK}
\author{R. Coldea}
\affiliation{
H.H. Wills Physics Laboratory, University of Bristol, Tyndall Avenue, 
Bristol, BS8 1TL, UK}
\author{T. S\"{o}rgel}
\author{M. Jansen}
\affiliation{
Max-Planck Institut f\"{u}r Festk\"{o}rperforschung, Heisenbergstrasse
1,
D-70569 Stuttgart, Germany
}

\date{\today}

\begin{abstract}
Zero-field muon-spin relaxation
 measurements of the frustrated  triangular quantum magnet AgNiO$_{2}$
are consistent with
a model of charge disproportionation that has been advanced to explain
the structural and magnetic properties of this compound.
Below an ordering temperature of $T_{\mathrm{N}}=19.9(2)$~K we observe 
six distinct muon precession frequencies, due to the magnetic order,
which can be accounted for with a model describing the probable
muon sites. 
The precession frequencies show an unusual temperature evolution
which is suggestive of 
the separate evolution 
of two opposing magnetic sublattices.
\end{abstract}

\maketitle

\begin{figure}
\begin{center}
\epsfig{file=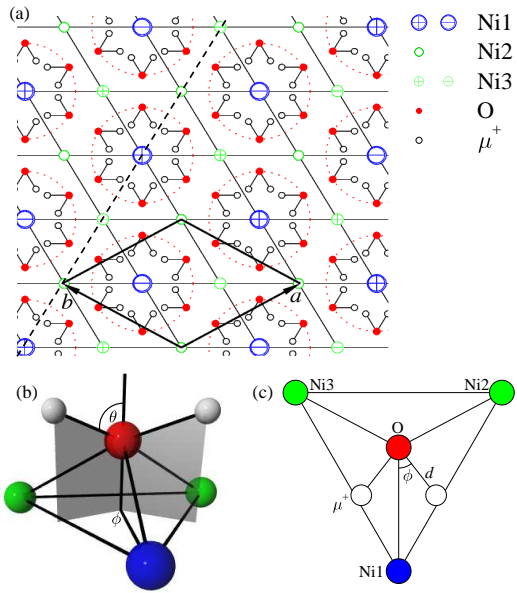,width=7cm}
\caption{(Color online) (a) Structure of AgNiO$_{2}$ viewed along the
$c$-direction with proposed muon stopping sites shown. The magnetic
structure involves ordered Ni1 sites with moments aligned parallel
($+$ sign) or antiparallel ($-$ sign) to the $c$-direction. 
surrounded by a honeycomb of Ni2 and Ni3 sites. The $+$ and $-$
signs on the Ni3 sites give the orientation of the Ni1 spins 
in the neighboring layer directly above and below.
The thick lines show the chemical unit cell, while the dashed line indicates
 the
mirror plane for the magnetic structure
(see text).
(b) Proposed muon stopping sites
occur 1~\AA\ from each oxygen ion.  Taking the oxygen position
as the origin of coordinates, muons are found at 
$\theta= 65^{\circ}$ and $\phi= \pm 38 ^{\circ}$, separated by
1.4~\AA\ from the Ni planes along $c$.
(c) As (b) but viewed along $c$. 
\label{muonsite}}
\end{center}
\end{figure}

Magnetic frustration of spins in triangular-lattice
antiferromagnets has been studied in a variety of experimental systems
\cite{collins,harrison}. The series $X$NiO$_{2}$, in which planes
of Ni$^{3+}$ 
($t_{2\mathrm{g}}^{6} e_{\mathrm{g}}^{1}$) ions
are arranged at the vertices of a triangular lattice,
provides the possibility that not only the magnetic, but also
charge and orbital degrees of freedom may be frustrated.
Although it is expected that the four-fold ground-state degeneracy of an 
isolated Ni$^{3+}$ ion would be lifted via a 
cooperative Jahn-Teller (JT) distortion 
(implying orbital ordering \cite{khomskii}) 
followed by magnetic ordering at a lower temperature \cite{vernay},
the triangular geometry of the $X$NiO$_{2}$ ($X$=Na, Li, Ag) system may 
frustrate both of these transitions and the ground state of such a 
system continues to be a matter of debate. \cite{vernay,mostovoy,chung}

Three energy scales determine the physics of these orbitally
degenerate systems \cite{mazin}:
the on-site Coulomb energy $U$, the electronic bandwidth $W$ and the
intra-atomic Hund's rule coupling $J_{\mathrm{H}}$. 
In insulating systems (such as NaNiO$_{2}$),  
for which $U \gg W$, the degeneracy of partially occupied orbital states 
is usually lifted by a cooperative JT distortion as described above. 
This contrasts
with an itinerant system for which $W \gg U$ (such as LaNiO$_{2}$) 
where the kinetic energy
term dominates and the JT distortion is suppressed. It has
recently been proposed \cite{mazin} that in an intermediate 
regime where $U$ and $W$ are comparable, $J_{\mathrm{H}}$ may overcome
the on-site repulsion leading to charge ordering (CO).
This provides another method for the
system to lift its orbital degeneracy; this time by removing
degenerate 
electrons at some sites, leading to double occupation at others. 

Such a CO scheme has been advanced \cite{radu} 
to explain the crystal structure and magnetic properties of
recently-synthesised hexagonal AgNiO$_{2}$,
%(space group P$6_{3}22$),
a polymorph \cite{sorgel,sorgel2} 
of the better known rhombohedral polytype \cite{shin}. This material
is a moderately delocalized metal \cite{sorgel} whose orbital degeneracy
may be lifted \cite{radu} by a 
$\sqrt{3} \times \sqrt{3}$ CO transition,
 achieved via $3 e_{g}^{1} \rightarrow e_{g}^{2} + 2 e_{g}^{0.5} $. 
This transition is indicated by a structural distortion
to a tripled unit cell with expanded and contracted NiO$_{6}$
octahedra. 
The $e_{g}^{2}$ ions, which correspond to $S=1$, Ni$^{2+}$
are located at ($\frac{1}{3}$, $\frac{2}{3}$, $\frac{1}{4}$) 
and these so-called Ni1 sites form the
tripled triangular lattice as shown in Fig~\ref{muonsite}(a).
The Ni1 spins order magnetically below $T=20$~K 
in a collinear structure of spins aligned along the $c$-direction,
organized in FM rows of stripes running
parallel to $b$ and alternating in direction along $a$ 
(Fig~\ref{muonsite}(a)). 
The $e_{g}^{0.5}$ sites (Ni2 located at ($0$, $0$,
$\frac{1}{4}$) and Ni3 at ($\frac{1}{3}$, $\frac{2}{3}$, $\frac{3}{4}$)), 
form a honeycomb network around the Ni1 sites and
appear to have a much reduced 
ordered moment ($\leq 0.1 \mu_{\mathrm{B}}$) due to hybridization \cite{radu}.
It should be noted that complete disproportionation is not required in
order to foster the complete spin polarization on a
particular site that is proposed in the model. The
delocalized nature of the system makes it possible
for the band of states corresponding to a class of site 
to be completely polarized,
despite being composed of a mixture of $e_{g}^{2}$ and $e_{g}^{0.5}$
states \cite{mazin}.

Muon-spin relaxation ($\mu^{+}$SR) measurements have been employed
to explore the properties of the $X$NiO$_{2}$ system
 for $X=$Na \cite{baker} and Li \cite{chatterji} from a local perspective.
In this paper we report $\mu^{+}$SR measurements made on
AgNiO$_{2}$. Our results may be explained in terms of the CO
model described above providing further evidence for the existence
of this newly proposed phase.
We show that the magnetically ordered
phase shows very little disorder, with six muon precession
frequencies observed with very little dephasing. 
However, the temperature evolution of the frequencies is unusual and points to 
the separate evolution of the Ni1 sublattice (described above) and an
additional sublattice that may be formed from small moments on the Ni2
and Ni3 sites.

Powder samples of hexagonal 
AgNiO$_{2}$ (with $< 1$\% of the rhombohedral 
polytype) were
prepared as described in Ref.~\onlinecite{sorgel}.
Zero field (ZF) $\mu^{+}$SR measurements were been made 
on the sample using the DOLLY instrument at the Swiss Muon Source (S$\mu$S).
The sample was wrapped in 25~$\mu$m Ag foil and mounted
on a Ag backing plate. 
In a $\mu^{+}$SR experiment \cite{steve}, spin-polarized
positive muons are stopped in a target sample, where the muon usually
occupies an interstitial position in the crystal.
The observed property in the experiment is the time evolution of the
muon spin polarization, the behavior of which depends on the
local magnetic field $B$ at
the muon site, and which is proportional to the
positron asymmetry function $A(t)$.

\begin{figure}
\begin{center}
\epsfig{file=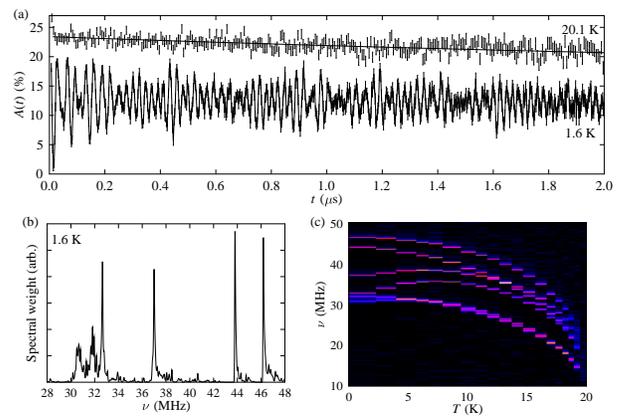,width=8.cm}
\caption{(Color online) (a) Time-domain $\mu^{+}$SR spectrum measured above
 and 
below the antiferromagnetic transition with time domain
fits as described in the main text. 
(b) Maximum entropy analysis of the spectrum measured at $T=1.6$~K
showing six distinct frequencies. (c) Temperature evolution of
the frequency spectrum extracted from maximum entropy analysis.
\label{data}}
\end{center}
\end{figure}

Typical ZF $\mu^{+}$SR spectra measured above and below
$T_{\mathrm{N}}$  
are shown in Fig.~\ref{data}(a). Above the magnetic transition the
spectra are well described by an exponential relaxation function as
expected for a paramagnet with dynamically fluctuating magnetic
moments \cite{hayano}.
Below $T_{\mathrm{N}}$
we observe oscillations in the time dependence of the muon
polarization (the ``asymmetry'' \cite{steve}) which are
characteristic of a quasi-static local magnetic field at the 
muon stopping site. This local field causes a coherent precession of the
spins of those muons for which a component of their spin polarization
lies perpendicular to this local field (expected to be $\frac{2}{3}$ of the
total spin polarization). 
The frequency of the oscillations is given by
$\nu_{i} = \gamma_{\mu} |B_{i}|/2 \pi$, where $\gamma_{\mu}$ is the muon
gyromagnetic ratio ($=2 \pi \times 135.5$~MHz T$^{-1}$) and $B_{i}$
is the average magnitude of the local magnetic field at the $i$th muon
site. Any fluctuation in magnitude of these local fields will
result in a relaxation of the oscillating signal, described by
relaxation rates $\lambda_{i}$ (see Eq.~(1)). 

The measured spectra for AgNiO$_{2}$ are unusual in that they
are composed of oscillations at several frequencies with very low
depolarization rates. This suggests the existence of several, well
defined, magnetically inequivalent muon sites in the material.
In order to account for all frequencies in the spectra measured below
$T_{\mathrm{N}}$, maximum entropy analysis was
used to decompose the signal. The results
are shown in Fig.~\ref{data}(b) and (c), where six well defined
frequencies are seen, corresponding to six magnetically
inequivalent muon sites in the material. We label these
$\nu_{1}$--$\nu_{6}$ in order of decreasing magnitude at $T=1.6$~K.
$\mu^{+}$SR spectra are recorded in the time domain and 
a knowledge of the number of frequencies allows us to use
time domain fitting for more detailed analysis. 
In a magnetically ordered 
powder with $N$ magnetically inequivalent muon sites, 
we would expect the spectra to be of the form
\begin{equation}
A(t) = \sum_{i=1}^{N} A_{i} \left( \frac{1}{3} + \frac{2}{3} 
  \mathrm{e}^{-\lambda_{i} t } \cos (2 \pi
\nu_{i} t + \phi_{i}) \right) +A_{\mathrm{bg}},\label{general}
\end{equation}
where $A_{\mathrm{bg}}$ represents a constant background contribution from those
muons that stop in the sample holder or cryostat tail and the
$\frac{1}{3}$ 
term
accounts for that fraction of muon-spin
components that lie parallel to the local magnetic field.
The measured spectra were fitted to Eq.~(\ref{general}), with $N=6$.
Constant nonzero phases $\phi_{i}$ were also required to fit the
data due to the difficulty in determining $t=0$ in the spectra
and the high oscillation frequencies involved. 
All components occurred with the same amplitude 
$A_{i} =  A(0)/6$ across the temperature range $T < T_{\mathrm{N}}$
from which we
conclude that each of the muon sites occurs with the same probability
and that the material is magnetically ordered throughout its bulk.

\begin{figure}
\begin{center}
\epsfig{file=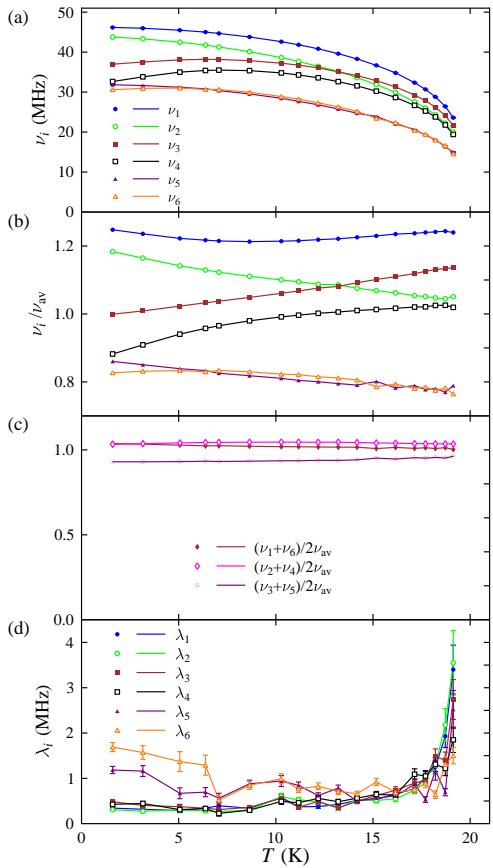,width=7.cm}
\caption{(Color online) 
(a) Temperature evolution of the frequencies $\nu_{i}$ extracted from 
the spectra measured below $T_{\mathrm{N}}$ fitted to
Eq.(\ref{general}). 
(b) The normalised frequencies $\nu_{i}/ 2 \nu_{\mathrm{av}}$
form three pairs with roughly equal and opposite gradients.
(c) Adding the pairs of normalised frequencies as indicated results
in approximately flat lines. 
(d) Relaxation rates $\lambda_{i}$ showing a steep increase as
$T_{\mathrm{N}}$ is approached from below.
\label{frequencies}}
\end{center}
\end{figure}

The temperature evolution of the fitted frequencies is shown in
Fig.~\ref{frequencies}(a), which is seen to be identical to that
extracted from the maximum entropy analysis [(Fig.\ref{data}(c))]. 
The magnitude of the frequencies $\nu_{i}$ 
are expected to act as an order parameter for a magnetic system.
The variation of these frequencies is therefore unusual, with some
frequencies decreasing and some increasing with increasing
$T$. 
However, the average frequency
$\nu_{\mathrm{av}}$ behaves as expected, decreasing smoothly
as $T$ is increased towards $T_{\mathrm{N}}$. Fits of
the average frequency to the phenomenological form 
$\nu_{\mathrm{av}}(T) =
\nu_{\mathrm{av}}(0)(1-(T/T_{\mathrm{N}})^{\alpha})^{\beta}$ yield
$\alpha \approx 2.4$, $\beta \approx 0.3$ and $T_{\mathrm{N}} =
19.9(2)$~K. This estimate for $T_{\mathrm{N}}$ is consistent with
the results of previous studies of AgNiO$_{2}$ \cite{sorgel}.

The relaxation rates $\lambda_{i}$ are
shown in Fig.~\ref{frequencies}(d). 
For $T <17$~K we note that the magnitudes of
all relaxation rates $\lambda_{i}$ are very low. 
The relaxation rates are expected to vary 
as $\lambda \sim \Delta^{2} \tau$, where
$\Delta$ is the second moment of the local magnetic field
distribution and $\tau$ is its fluctuation time \cite{hayano}. 
This implies
that the distribution of local fields is exceptionally narrow at all muon
sites and that the spin disorder in AgNiO$_{2}$ is remarkably low. 
All $\lambda_{i}$ have very similar temperature dependence
except $\lambda_{5}$ and $\lambda_{6}$, which are larger than
$\lambda_{1-4}$ across the entire temperature regime and show a rough
decrease with increasing temperature for $T < 17$~K.
As $T_{\mathrm{N}}$ is approached from below, all $\lambda_{i}$ 
increase sharply as would be expected
for the onset of critical fluctuations near a magnetic transition. 

A knowledge of $\nu_{i}(T=0)$ allows a determination of the muon sites
in AgNiO$_{2}$. Dipole fields were 
calculated in a sphere containing $\sim 10^5$ Ni ions, using the 
magnetic structure described above \cite{radu}. 
A positive muon's stopping position is usually 1~\AA\ from the electronegative 
O$^{2-}$ ion \cite{holzschuh} and calculated local magnetic fields corresponding
to the observed frequencies are found
at several positions around 1~\AA\ from each oxygen ion. 
In order to narrow the choice of muon sites, the electrostatic potential was 
also calculated at positions on the 1~\AA\ sphere surrounding
 an oxygen, assuming the proposed \cite{radu} $\sqrt{3} \times \sqrt{3}$ 
CO on the Ni sites. 
Two distinct regions of negative electrostatic potential are found,
separated from the Ni planes along $c$ by 1.4~\AA\ and
occurring  closer to the Ni1 ions (which have $e_{g}^{2}$) 
than the Ni2 and Ni3 ions (with $e_{g}^{0.5}$). These regions 
are found to contain the positions where the local
magnetic fields give rise to the observed precession frequencies.
For each oxygen ion we may therefore identify two muon sites.
We take an oxygen as the origin of a polar coordinate system 
(shown in Fig.~\ref{muonsite}(b)) with
$z$-direction parallel to the $c$-axis and
$x$-direction defined by the projection onto the $a$--$b$ plane
of the vector joining the O$^{2-}$ and the Ni1 ions. The muon
sites are found 1~\AA\ from the origin at $\theta=\pm 65^{\circ}$
(plus sign for oxygens above the Ni planes, minus for those
below) and $\phi= 38 ^{\circ}$ and $\phi= -38 ^{\circ}$. These
proposed muon sites are shown in Fig~\ref{muonsite}.
For the twelve muon sites surrounding a Ni1 ion (Fig.~\ref{muonsite}(a)), 
the proposed magnetic structure involving
alternating ferromagnetic rows of ordered
moments gives rise to six magnetically distinct sites.
This is because the magnetic structure has a mirror plane that
runs transverse to the magnetic stripe
direction (Fig.~\ref{muonsite}(a)) and which causes sites linked by
this symmetry to be equivalent.
This approach accounts successfully for the six observed magnetic 
frequencies. Unfortunately, however, the large field gradients
in these positions, along with the possibility of small moments on the Ni2 and
Ni3 sites, make it difficult to assign precise positions for each of the
observed frequencies. 

We now turn to the unusual temperature dependence of the frequency
spectrum. 
The exceptionally small relaxation rates permit the extraction of
the detailed temperature dependence of the local magnetic
field at each muon site. 
Normalizing the frequencies $\nu_{i}$ by the average value 
$\nu_{i}/2\nu_{\mathrm{av}}$
(Fig.~\ref{frequencies}(b)) 
gives
three pairs with
approximately equal and opposite gradients. These pairs
($\nu_{1}+\nu_{6}, \nu_{2}+\nu_{4}$ and $\nu_{3}+\nu_{5}$)
are shown in Fig.~\ref{frequencies}(c) to yield approximately 
flat lines when added together, demonstrating that they vary in equal 
and opposite
senses about the average.  This suggests that within a pair, one muon
site sees an increase in local magnetic field at the expense of
its partner, which sees a decrease, as temperature is increased. 
In a polycrystalline material the muon-spin polarization depends only
on the {\it magnitude} of the local magnetic field at the muon site.
Our picture of the six muon sites in AgNiO$_{2}$
therefore comprises three pairs of sites, where the pairs see opposing
variation with temperature in the magnitude of the local dipole fields. 
Fig.~\ref{frequencies}(b) shows that this effect is quite small (but
clearly detectable)
and that
the variation in local field
at a particular muon site is less than 
$\lesssim 15$\%  about the average.

A possible explanation for this anomalous behavior involves
a continuous temperature evolution of the direction
of the ordered moment, so that the field magnitude increases at one muon
site whilst correspondingly being reduced at another. In a
polycrystalline material, such a reorientation would
not change the amplitudes associated with each frequency component
due to the effects of angular averaging, so would be consistent
with our measurements. However, it is difficult to see how a
reorientation would be energetically favoured as it would involve
the interactions between spin components in the $c$-direction
(i.e. along the local 3-fold easy axis)
driving a reorientation directions in the $a$-$b$ plane
along a further 3-fold symmetric easy axis. If these
components were to order with the same wavevector as the components
aligned along $c$, two of the three in-plane spins could not point
along their local easy axis direction. 
We note further that it is unlikely that this effect is caused by a 
temperature dependent reordering of charges. 
The structural distortion that indicates the CO transition occurs
above room temperature and involves the expansion and contraction
of NiO$_{6}$ octahedra. 
Any charge reordering should be observable as a change on Ni--O bond
lengths, which is not seen in neutron diffraction studies \cite{wawrzynska}

A more probable scenario involves the separate temperature evolution of
two magnetic sublattices. 
A second magnetic sublattice could arise from Ni2 and
Ni3 sites, which band-structure calculations
suggest \cite{radu} may carry a magnetic moment of
$0.1~\mu_{\mathrm{B}}$. An ordered moment on this second sublattice
may be induced by the large ordered moment
on the Ni1 sites and would, therefore, presumably lie along $c$.
Although the temperature
dependence of the Ni1 moment will be determined by Ni1
intra-sublattice exchange, the $T$ dependence
of the Ni2,3 sublattice would most likely depend not only on its
intra-sublattice exchange but also on its interaction with the 
large Ni1 moment,
possibly causing the two sublattices to evolve differently in temperature.
This scenario would
require that the pairs
of muon sites be arranged such that a decrease at one site would
also lead to an increase at another. 
For example, if the Ni3 sites become spin-polarized by interaction
with the Ni1 sites in the planes immediately above and below, the Ni3
sublattice will give rise to a small additional dipolar field which
augments that produced by the Ni1 sublattice for half the muon sites,
but
diminishes the field for the other half. Thus a temperature-dependent
interplanar coupling between Ni1 and Ni3 sites could give rise to the
anomalous temperature evolution that we have observed.

This work was carried out at S$\mu$S, 
Paul Scherrer Institute, Villigen, Switzerland. 
We are grateful to Robert Scheuermann for experimental
assistance. This work is supported by the EPSRC. 
T.L. acknowledges support from the Royal Commission for the Exhibition
of 1851.


\begin{thebibliography}{99}

\bibitem{collins}
M.F. Collins and O.A. Petrenko, Can. J. Phys. {\bf 75}, 605 (1997).

\bibitem{harrison}
A. Harrison, J. Phys. Condens. Matter {\bf 16}, S553 (2004).

\bibitem{khomskii}
D.I. Khomskii, Physica Scripta {\bf 72}, 8 (2005).

\bibitem{vernay}
F. Vernay {\it et al.}, 
%K.Penc, P.Fazekas and F. Mila, 
Phys. Rev. B {\bf 70},
014428 (2004).

\bibitem{mostovoy}
M.V. Mostovoy and D. I. Khomskii, Phys. Rev. Lett. {\bf 89}
227203 (2002).

\bibitem{chung}
J.-H. Chung {\it et al.}, 
%Th. Proffen, S. Shamoto, A.M. Ghorayeb, L. Corguennec, 
%W. Tian, B.C. Sales, R. Jin, D. Mandrus and T. Egami
Phys. Rev. B, {\bf 71} 064410 (2005).

\bibitem{mazin}
I.I. Mazin {\it et al.}, 
%D.I. Khomskii, R. Lengsdorf, J.A. Alonso, W.G. Marshall, 
%R.M. Ibberson, A. Podlesnyak, M.J. Mart\'{i}nez-Lope and
%M.M. Abd-Elmeguid, 
Phys. Rev. Lett. {\bf 98} 176406 (2007).

\bibitem{radu}
E. Wawrzy\'{n}ska {\it et al.}, 
%R. Coldea, E.M. Wheeler, I.I. Mazin, M.D. Johannes,
%T. S\"{o}rgel, M. Jansen, R.M. Ibberson and P.G. Radaelli,
arXiv:0705.0668v2. 

\bibitem{sorgel}
T. S\"{o}rgel and M. Jansen, Z. Anorg. Allg. Chem. {\bf 631}, 2970
(2005).

\bibitem{sorgel2}
T. S\"{o}rgel and M. Jansen, J. Solid State Chem. {\bf 180}, 8
(2007).

\bibitem{shin}
Y.J. Shin {\it et al.}, 
%J.P. Doumerc, P. Dordor, C. Delmas, M. Pouchard
%and P. Hadenmuller, 
J. Solid. State Chem. {\bf 107}, 303 (1993).

\bibitem{baker}
P.J. Baker {\it et al.}, 
%T. Lancaster, S.J. Blundell, M.L. Brooks, W. Hayes, 
%D. Prabhakaran and F.L. Pratt, 
Phys. Rev. B, {\bf 72}, 104414 (2005).

\bibitem{chatterji}
T. Chatterji, W. Henggeler and C. Delmas, J. Phys. Condens. Matter
{\bf 17} 1341 (2005).

\bibitem{steve}
S.J. Blundell, Contemp. Phys. {\bf 40}, 175 (1999).


\bibitem{hayano}
R.S. Hayano {\it et al.}, 
%Y. J. Uemura, J. Imazato, N. Nishida, T. Yamazaki and R. Kubo,
 Phys. Rev. B {\bf 20}, 850 (1979).

\bibitem{holzschuh}
E. Holzschuh {\it et al.}, 
%A. B. Denison, W. K\"{u}ndig, P.F. Meier and
%B.D. Patterson, 
Phys. Rev. B {\bf 27}, 5294 (1983).

\bibitem{wawrzynska}
E. Wawrzy\'{n}ska {\it et al.}, 
%R. Coldea, E.M. Wheeler, T. S\"{o}rgel, M. Jansen, 
%R. Ibberson, P.G. Radaelli and M. Koza, 
unpublished. 

\end{thebibliography}
\end{document}